\documentclass[prl,twocolumn,amsmath,amsfonts,amssymb,showpacs]{revtex4} 
\usepackage{epsfig,psfrag,graphicx,bm,color}
\begin{document}
\title{New Positron Spectral Features from Supersymmetric Dark Matter\\ -- a Way to Explain the PAMELA Data?}

\author{Lars Bergstr\"om}
\email{lbe@physto.se}
\author{Torsten Bringmann}
\email{troms@physto.se}
\author{Joakim Edsj\"o}
\email{edsjo@physto.se}
\affiliation{Department of Physics, Stockholm University, AlbaNova
 University Center, SE - 106 91 Stockholm, Sweden}

\date{August 27, 2008}

\pacs{95.35.+d,98.70.Sa,13.40.Ks, 11.30.Pb}

\begin{abstract}
The space-borne antimatter experiment PAMELA has recently reported a surprising
 rise in the positron to electron ratio at high energies. It has also 
recently been  found 
that electromagnetic radiative corrections in some cases may boost 
the gamma-ray yield from supersymmetric dark matter annihilations in the 
galactic halo by up to three or four orders of magnitude, providing distinct 
spectral signatures for indirect dark matter searches to look for. Here, we 
investigate whether the same type of corrections can also lead to sizeable 
enhancements in the positron yield. We find that this is indeed the case, 
albeit for a smaller region of parameter space than for gamma rays; selecting
models with a small mass difference between the neutralino and sleptons,
like in the stau coannihilation region in mSUGRA, the effect becomes more pronounced. The resulting, rather hard positron 
spectrum with a relatively sharp cutoff may potentially fit the 
rising positron ratio measured by the PAMELA satellite. To do so, however,
very large ``boost factors'' have to be invoked that are not expected in 
current models of halo structure. If the predicted 
cutoff would also be confirmed by later PAMELA data or upcoming
 experiments, one could either assume non-thermal production in 
the early universe 
or non-standard halo 
formation to explain such a spectral feature as an effect 
of dark matter annihilation. 
At the end of the paper, we briefly comment on the impact of radiative 
corrections on other 
annihilation channels, in particular antiprotons and neutrinos.

\end{abstract}

\maketitle

\newcommand{\ga}{\gamma}
\newcommand{\be}{\begin{equation}}
\newcommand{\ee}{\end{equation}}
\newcommand{\bea}{\begin{eqnarray}}
\newcommand{\eea}{\end{eqnarray}}
\newcommand{\ds}{{\sf DarkSUSY}}
\newcommand{\py}{{\sf PYTHIA}}
\newcommand{\code}[1]{{\tt #1}}

\hyphenation{}

The existence of a sizeable dark matter contribution to the total cosmological 
energy density seems by now to be established beyond any reasonable doubt, the 
most recent estimates \cite{wmap5} giving the fraction of cold dark 
matter to 
the critical density as $\Omega_{CDM}\sim 0.233 \pm 0.013$.  
Against this 
background, searches for experimental signatures that may determine the so far 
still elusive nature of the cosmological dark matter are becoming ever more 
important.

On the theoretical side,  maybe the best motivated, and certainly 
most extensively studied, dark matter candidate is the supersymmetric 
neutralino  (for reviews, see \cite{reviews}). The methods of detection for 
this type of particle dark matter can be grouped into {\em accelerator} 
searches, trying to directly produce dark matter or related new particles  (the 
signature of the former usually being missing 
energy), the {\em direct  
detection} of dark matter particles scattering off the nuclei of a 
terrestrial 
detector, or {\em indirect 
detection} of particles
generated by the 
annihilation of dark matter particles in the Galactic halo or (for neutrinos) 
in the Sun or Earth.
With the LHC soon
operating and new detectors of liquid noble 
gases  being developed for 
direct detection, aiming to further improve the 
already impressive recent upper limits \cite{direct}, the near future promises 
very interesting times for the field. 

As far as indirect detection is 
concerned, the antimatter detection satellite PAMELA 
has just announced its first set of data for cosmic ray 
antiprotons \cite{pameladata} and has very recently done so also for 
positrons \cite{pamela_idm08}. Although the antiproton data seems to agree with 
conventional secondary production by cosmic rays, the positron data shows 
an unexpected rise in the differential ratio $e^+/(e^-+e^+)$ above some 7 -- 10 GeV. This interesting situation may be further investigated by the PEBS balloon
experiment \cite{pebs} and, in particular, the AMS-02 experiment, if installed on the international space 
station \cite{AMS-02}. These experiments could further improve these data, 
both concerning 
statistics and energy range, and in particular investigate whether a return to a ``normal'' ratio at some energy exists -- something that is predicted by models
of dark matter annihilation due to the kinematic limit that appears at an energy equal 
to the dark matter particle mass. Of course, it will also be important to 
rule out positron misidentification through proton contamination in this 
high-energy range.  

Further information may possibly be obtained from  the huge 
IceCUBE \cite{icecube} detector which will soon start to 
look for cosmic neutrinos at the South Pole. For gamma-rays, 
the recently
launched GLAST satellite \cite{glast} opens up a
 new window to the 
high-energy
universe, for energies from below a GeV to about 300 GeV. The 
sensitivity to gamma-rays of even higher energies is, furthermore, expected to 
improve considerably with next generation Air Cherenkov Telescopes like the CTA 
or AGIS \cite{cta,agis}. 

Most likely, in fact,  a signal 
from more than one 
type of experiment will be needed to confirm a dark matter interpretation of the observed signal and, in the best case, to
fully identify the 
particle making up the 
dark matter; 
it is thus important to realize the complementary nature of the 
methods described above. This is even more true since, in all of these cases, 
the signal searched for 
may be 
quite weak and dominated by a much larger 
background. In this context, one should also keep in mind (see, e.g., \cite{atoyan95}) that it may be possible to explain a signal like the one recently reported by PAMELA in a more conventional way, without having to invoke unreasonably strong astrophysical sources; for example, 
a supernova remnant of age $10^5$ years some 100 pc distance from the Sun/Earth
would both have the appropriate energetics and the right energy spectrum
to account for the PAMELA results. 

A first assessment of the situation (see, e.g., \cite{gondolo_idm08}) concerning dark matter candidates after
the surprising PAMELA results 
seems to indicate
that the otherwise favoured supersymmetric neutralino 
cannot 
explain the data. This is because it is a Majorana particle and therefore does
not 
give hard positrons directly, due to the helicity suppression of light 
fermions in the annihilation process. The resulting positron spectrum is thus expected to be rather soft, in disagreement with the PAMELA data,
and therefore a Dirac particle, or a spin-1 particle  like Kaluza-Klein 
dark matter \cite{Hooper:2004xn} would fit better (another proposal put forward in connection with the PAMELA data has been minimal dark matter, where the combination of a very high dark matter particle mass ($\sim10$~TeV) and a very efficient enhancement mechanism for the annihilation into charged gauge bosons would result in the required hard spectrum at low energies \cite{mdm}). 

However, this simple intuition may prove wrong when computing radiative 
corrections. Gamma-rays, for instance, have a sharp cutoff 
\cite{Bergstrom:2004cy,birkedal} at an energy equal to 
the dark matter particle's mass, $E_\gamma=m_\chi$, and, in some cases, even 
prominent line signals from the direct annihilation into photons 
\cite{lines,hisano,idm}. While the origin of the first feature is associated 
with photons directly radiated from charged final legs (``final state 
radiation''), it was recently pointed out that even photons radiated from 
charged virtual particles (``virtual'' internal bremsstrahlung (IB), or direct emission) can 
have a significant impact on the resulting gamma-ray spectrum, leading not only 
to an even more pronounced cutoff, but also to clearly observable bump-like 
features at slightly lower energies \cite{ibsusy}. In fact, these effects 
generically dominate the total spectrum at high photon energies, including even 
the line signals, and may lead to an enhancement of the annihilation rate by several orders of magnitude. Such a large radiative ``correction'' can 
appear since the annihilation of neutralinos into lepton pairs is strongly 
helicity suppressed, while for the three-body final state containing an 
additional photon this suppression is circumvented \cite{lbe89}.

With these 
recent results in mind, the question thus naturally arises whether the same 
effects also have a significant impact on, e.g., the yield in positrons -- 
especially since the largest enhancement factors appear for neutralino 
annihilations into leptons \cite{ibsusy}. 
Let 
us first consider the \emph{direct annihilation} into positrons. As $e^+e^-$ two-body
final states are strongly suppressed, the dominant contribution comes always from the 
process $\chi\chi\rightarrow e^+e^-\gamma$, in particular from those diagrams 
where the photon is radiated from a $t$-channel selectron. Setting 
$m_e\rightarrow 0$, we find for the differential annihilation rate into positrons:

\bea
\label{dNdx}
  &&\!\!\frac{d}{dx}(v\sigma)^{\chi\chi\rightarrow e^+e^-\gamma}_{v\rightarrow0} = \frac{\alpha_\mathrm{em}\left|\tilde g_R\right|^4}{256\pi^2m_\chi^2}\frac{1}{({\mu_R}\!-\!1\!+\!2x)^2}\\
&&\times\Bigg\{\!\!\left[4(1\!-\!x)^2-4x(1\!+\!{\mu_R})+3(1\!+\!{\mu_R})^2\right]\log\frac{1\!+\!{\mu_R}}{1\!+\!{\mu_R}\!-\!2x}\nonumber\\
&&\phantom{\times\Big\{}-\left[4(1\!-\!x)^2-x(1\!+\!{\mu_R})+3(1\!+\!{\mu_R})^2\right]\frac{2x}{1\!+\!{\mu_R}}\Bigg\}\nonumber\\
&&+\Big(R\leftrightarrow L\Big)\,,\nonumber
\eea
where $x=E_{e^+}/m_\chi$, ${\mu_{R,L}}\equiv m_{{\tilde e}_{R,L}}^2/m_\chi^2$ 
and $\tilde g_RP_L$ ($\tilde g_LP_R$) is the coupling between neutralino, 
electron and right-handed (left-handed) selectron. 
In the corresponding limit, this 
reproduces the result found in \cite{lbe89} for photino annihilation. Integrating Eq.~(\ref{dNdx}) gives
\bea
\label{sigtot}
  &&\!\!(v\sigma)^{\chi\chi\rightarrow e^+e^-\gamma}_{v\rightarrow0} = \frac{\alpha_\mathrm{em}\left|\tilde g_R\right|^4}{64\pi^2m_\chi^2}\\
&&~\times\Bigg\{\frac{3\!+\!4{\mu_R}}{1\!+\!{\mu_R}}+\frac{4{\mu_R}^2\!-\!3{\mu_R}\!-\!1}{2{\mu_R}}\log\frac{{\mu_R}\!-\!1}{{\mu_R}\!+\!1}\nonumber\\
&&~\phantom{\times\Big\{}-\left(1\!+\!{\mu_R}\right)\!\left[\frac{\pi^2}{6}-\!\left(\!\log\frac{{\mu_R}\!+\!1}{2{\mu_R}}\right)^2\!-2\mathrm{Li}_2\!\left(\frac{{\mu_R}\!+\!1}{2{\mu_R}}\right)\right]\!\!\Bigg\}\nonumber\\
&&~+\Big(R\leftrightarrow L\Big)\,,\nonumber
\eea
where $\mathrm{Li}_2(z)=\sum_{k=1}^\infty z^k/k^2$ is the dilogarithm. The direct annihilation of neutralinos (with small galactic velocities $v$) into positrons is thus only suppressed by a factor of $\left(\alpha/\pi\right)$ and not, as often quoted, by the much smaller factor of $\left(m_e^2/m_\chi^2\right)$ that is connected with two-body final states. In the above expression, the highest annihilation rate is obtained in the limit $\mu_{R,L}\rightarrow1$ where the selectrons are degenerate in mass with the 
neutralino:
\bea
\label{sigtotmax}
  &&(v\sigma)^{\chi\chi\rightarrow e^+e^-\gamma}_{v\rightarrow0} \leq \frac{\alpha_\mathrm{em}}{\pi}\frac{\left|\tilde g_R\right|^4+\left|\tilde g_L\right|^4}{\pi m_\chi^2}\frac{21-2\pi^2}{384}.
\eea

Positrons may also be produced in the decay of other annihilation 
products. The number $dN_{e^+}^f/dx$ of such \emph{secondary} positrons per 
annihilation into the corresponding final state $f$ can be simulated with Monte 
Carlo event generators like \py\ \cite{PYTHIA}. For two-body final states 
$X\bar X$, we use the tabulated values contained in \ds\ \cite{ds} that were 
obtained through a large number of \py\ runs. For three-body final states 
containing a photon, the positron yield is approximately given by
\be
  \label{dNsec}
  \frac{dN_{e^+}^{X\bar X\gamma}}{dx}\approx\int dE_X \frac{dN_X^{X\bar X\gamma}}{dE_X}\frac{d\tilde N_{e^+}^{X\bar X}}{dx}\,,
\ee 
where ${d\tilde N_{e^+}^{X\bar X}/dx}$ is the (two-body final state) positron 
multiplicity  $dN_{e^+}^{X\bar X}/dx$ that results from the annihilation of two 
dark matter particles with mass $E_X$.  The analytically obtained expressions for $dN_X^{X\bar X\gamma}/dE_X$ are too lengthy to reproduce here but have been fully implemented in the current public release of \ds\ \cite{ds} (for light fermions, of course, the same functional form as in Eq.~(\ref{dNdx}) is recovered).
When compared to gamma rays from the corresponding channel, the 
positron contribution (\ref{dNsec}) to the total spectrum is considerably 
less pronounced at the observationally most relevant energies near the cutoff 
since part of the energy is taken away by the photon; the fact that positrons 
are not the only decay products induces a further kinematical suppression at 
high energies. On general grounds, we therefore cannot expect large radiative 
corrections to the yield in secondary positrons -- even in situations where 
large gamma-ray contributions are found (as, e.g., for heavy neutralino 
annihilation  into $W^+W^-$ \cite{heavysusy}). An exception to this conclusion 
could only occur in a situation where the annihilation rate into the three-body 
final state is many times larger than for the two-body final state. As pointed 
out in \cite{ibsusy}, this  is indeed possible for lepton final states in the 
stau-coannihilation region of mSUGRA. However, as also the annihilation into 
$e^+e^-\gamma$ is usually greatly enhanced in this region, it is, rather, the 
latter contribution that dominates in this case.

\begin{figure}[t!]
   \includegraphics[width=\columnwidth]{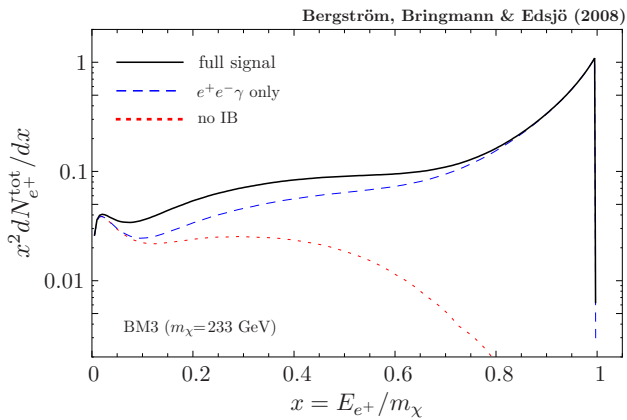}
\caption{The solid line gives the total number of positrons per neutralino pair 
annihilation and positron energy for the benchmark model BM3 of \cite{ibsusy} ($m_\chi=233$~GeV, $m_{\tilde e}=240$~GeV).  
Shown separately is the same quantity without radiative corrections (dotted 
line) and, on top of this, only the $e^+e^-\gamma$ final states (dashed line). 
\label{fig_dNdx}}
\end{figure}

For illustration, we show 
in Fig.~\ref{fig_dNdx} the effect of radiative 
corrections on the positron yield for a typical model in the mSUGRA coannihilation region (introduced as benchmark point BM3 in \cite{ibsusy}), which is characterized by small mass differences between neutralino and sleptons.
A spectacular boost in the positron yield can be observed, leading to an 
extremely pronounced cutoff at $E_{e^+}=m_\chi$. As anticipated, this is mainly 
due to primary positrons, following the distribution (\ref{dNdx}), but at 
smaller energies the effect of radiative corrections becomes also visible for 
other decay channels (mainly $\mu^+\mu^-$).

\begin{figure*}[t!]
   \includegraphics[width=0.9\columnwidth]{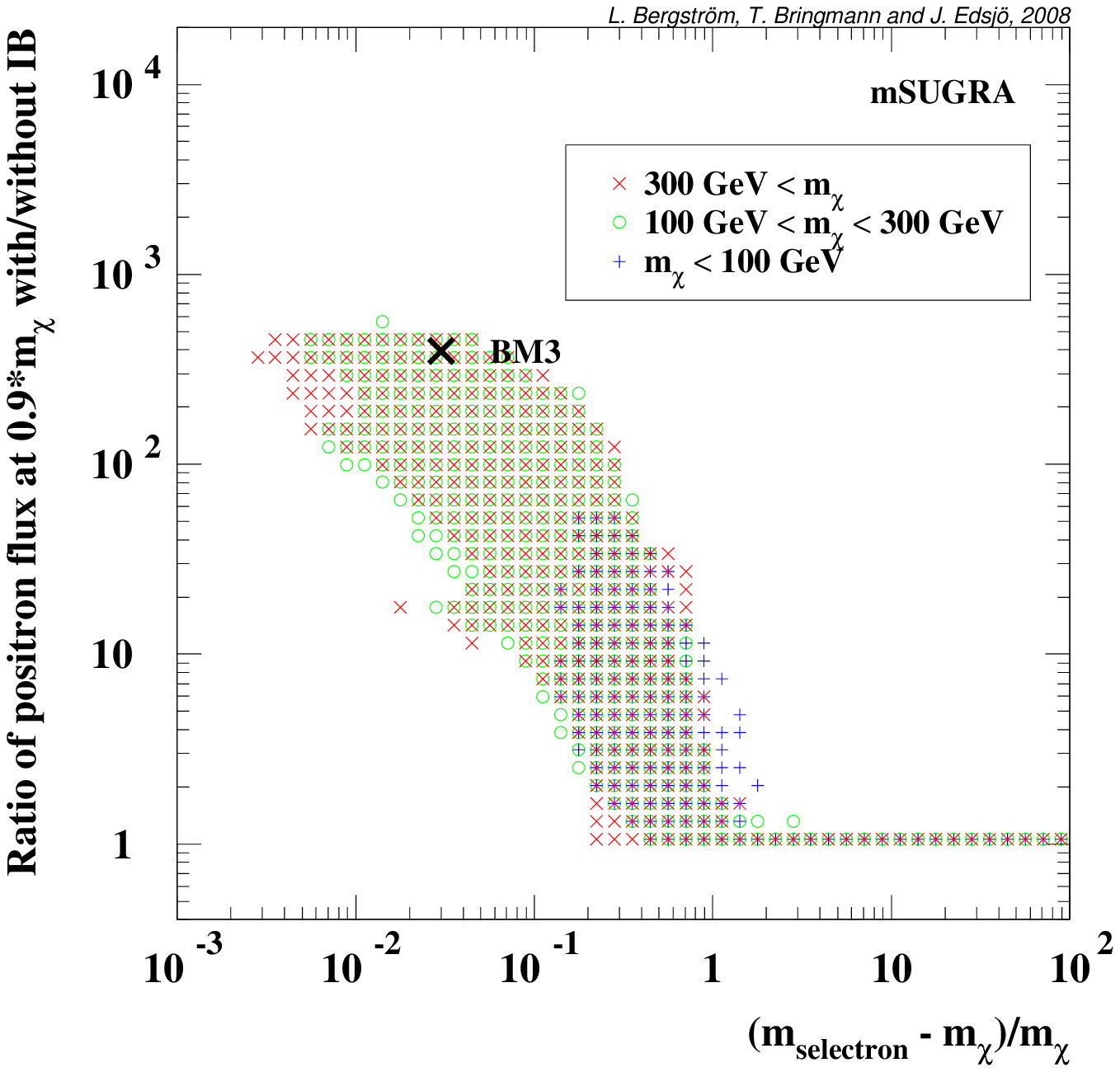}
   \includegraphics[width=0.9\columnwidth]{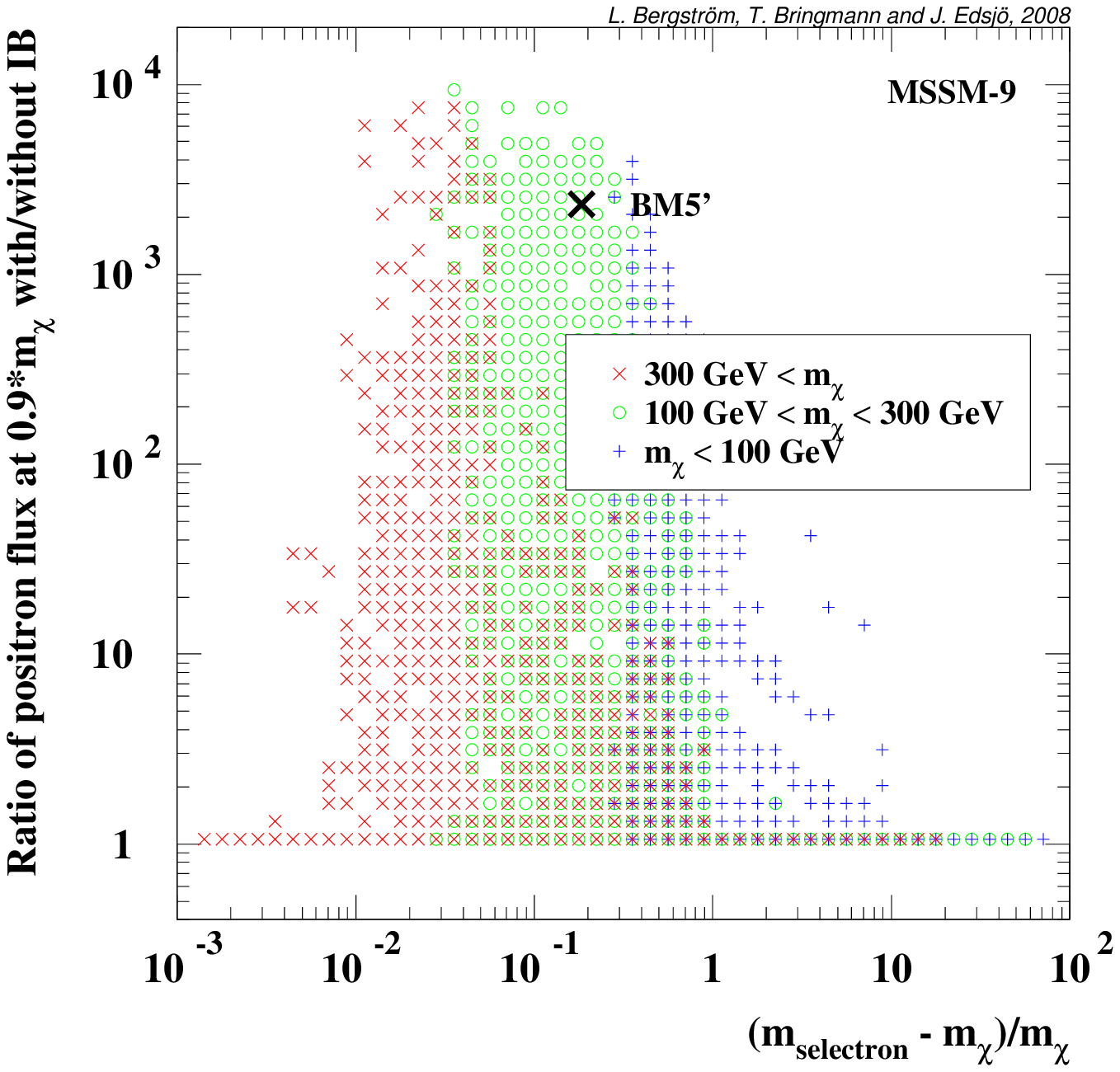}
\caption{Scan over mSUGRA (left) and MSSM-9 (right) models that shows the enhancement in the positron flux (at $E_{e^+}=0.9\,m\chi$)
due to radiative corrections  vs.~the 
mass splitting between the lightest selectron and the neutralino, 
$\delta\equiv\left(m_{\tilde e}-m_\chi\right)/m_\chi$. 
Also indicated in this 
figure are the benchmark model BM3 from \cite{ibsusy} and a further benchmark model BM5' as introduced in the text.
\label{fig_scatter}}
\end{figure*}

The propagation of charged particles is influenced by 
magnetic fields residing in the Milky Way which, in contrast to the case of photons, tend to erase clearly pronounced spectral features. To be able to compare our results with the cosmic ray positron 
spectrum as measured at the top of the atmosphere (TOA) or, in the case of the recent PAMELA data, in space, we adopt the standard 
assumption of randomly distributed magnetic fields, in which case the 
determination of the positron flux at a given galactic position boils down to 
solving a diffusion equation (for more details on the procedure we follow, see 
\cite{Baltz:1998xv}). We assume an NFW profile \cite{nfw} for the dark matter 
distribution in the galactic halo, but allow for an additional ``boost factor'' to 
account for the effect of dark matter substructure. 

Let us now quantify our general expectations outlined above and try to assess the general importance of IB effects on the positron yield. For that purpose, we consider a scan (based on the work in \cite{baltz_peskin})
over the mSUGRA parameter space, keeping only models that feature the right relic density and pass all current collider bounds. In this setup, as mentioned before, the stau coannihilation region is characterized by light leptons; since, at the same time, the total annihilation cross section (today, i.e.~for $v\rightarrow0$) is very small, we expect rather large enhancements in the positron yield in this case. In the left panel of Fig.~\ref{fig_scatter}, indeed, we clearly observe the expected strong correlation between the $m_{\tilde e}$-$m_\chi$ mass splitting and the resulting enhancement in the positron flux due to radiative corrections; outside the coannihilation region, no sizeable flux enhancements are encountered. 

In mSUGRA,
however, in the presence of light selectrons, also the other sleptons have to be light. Hence, to
investigate the situation in more general terms, we have set up a low-energy phenomenological scan with 9 free parameters.
We start with the usual parameters of MSSM-7: the Higgsino mass parameter $\mu$, the gaugino mass parameter $M_2$, a common
sfermion mass scale $m_0$, the ratio of the Higgs vacuum expectation values $\tan \beta$, the trilinear couplings in the third generation $A_t$ and $A_b$, and
the mass of the CP-odd Higgs boson, $m_A$. In addition to these, we add  a selectron mass
parameter that goes into the selectron entries in the mass parameters of the soft SUSY-breaking Lagrangian; furthermore, we
relax the GUT condition on $M_1$ and $M_2$ by allowing $M_1$ to be varied freely. We then generate models by
varying these 9 parameters of our MSSM-9 model between generous bounds, focusing on models with light selectrons and, again, keeping only models that feature the right relic density and pass all current collider bounds.
In the right panel of Fig.~\ref{fig_scatter}, we plot the enhancements from IB for these models; the effect of introducing more free parameters as compared to mSUGRA is mainly that the total annihilation cross section is not anymore closely linked to the $m_{\tilde e}$-$m_\chi$ mass splitting by the relic density requirement. As a result, the enhancements in the total positron flux can be both considerably larger and smaller than in the mSUGRA case, depending on the total annihilation rate to lowest order; the main contribution to the flux enhancement, however, is in any case found from the $e^+e^-\gamma$ channel.

\begin{figure}[t!]
   \includegraphics[width=0.95\columnwidth]{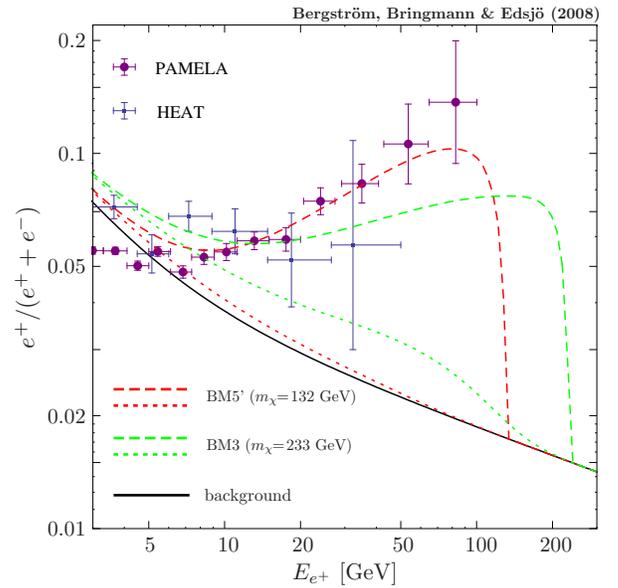}
\caption{The solid line is the expected flux ratio $e^+/(e^++e^-)$ as 
calculated following \cite{galprop}. The data points are the combined HEAT \cite{heat} and PAMELA data \cite{pamela_idm08}. Furthermore, the expected flux ratio for our benchmark models is shown without (dotted lines) and after taking into account radiative corrections (dashed lines). See text for further details.
\label{fig_flux}}
\end{figure}

In Fig.~\ref{fig_flux}, we plot  the resulting flux ratio  $e^+/(e^++e^-)$  from 
neutralino annihilations for both BM3 and a point BM5' in the MSSM-9 parameter space (with $m_\chi=132$~GeV, $m_{\tilde e}=157$~GeV; both models are marked in Fig.~\ref{fig_scatter}) and compare it to the PAMELA data. For comparison, we also show the 
expected background flux \cite{galprop}. Propagation effects thus 
considerably smear the 
spectrum shown in Fig.~\ref{fig_dNdx}, but the clearly pronounced cutoff at 
$E_{e^+}=m_\chi$ still remains as a prominent feature. 
It is interesting to 
note that this type of pronounced spectral signature so far has only been 
associated to Kaluza-Klein dark matter \cite{kkpos}. Even though the cutoff in 
this latter case appears, due to the large branching ratio into $e^+e^-$, to be even 
more pronounced, it would be observationally very challenging to see this 
difference with an energy resolution of the about 5\% expected for PAMELA. The apparent discrepancy between the background expectation and the new data at small energies is most likely due to a change in the solar potential which has not been taken into account so far \cite{pamela_idm08}; this effect, however, is expected to be negligible at positron energies above around 10 GeV and we will therefore not discuss it further here.

Unfortunately, the need for large boost factors is generic for all models that 
show a high positron yield enhancement in the way reported here (in 
Fig.~\ref{fig_flux}, we used boost factors of $3\times10^4$); the reason being that models in the coannihilation region have 
very small annihilation rates to start with. While such 
large boost factors are difficult to achieve in standard scenarios \cite{diemand}, they are 
easily encountered if, e.g., the Milky Way hosts a typical population of 
intermediate mass black holes \cite{Brun:2007tn}; other possibilities include a non-thermal neutralino production in the early universe (see, e.g., \cite{Kane:2002nm}), a non-standard pre-BBN expansion rate \cite{Arbey:2008kv}, or a very nearby dark matter ``clump'' (which, however, is quite unlikely according to present models of structure formation).

One should also keep in mind that boost factors 
of at least 50-100 are needed in most cases, anyway  \cite{Baltz:2001ir}, 
to see the effect of supersymmetric 
dark matter annihilation in the positron spectrum -- but even for very large boosts, the resulting positron spectra are too soft to explain the observed steep rise in the $e^+/(e^++e^-)$ ratio.
As becomes apparent from Fig.~\ref{fig_flux}, this is actually also true for the already quite hard spectra reported here in case of masses higher than $m_\chi\gtrsim100$~GeV. In order to really fit the PAMELA data through primary positrons from neutralino annihilations would thus require rather small neutralino masses. A generic prediction of this model is therefore that a sharp cutoff in the spectrum has to be observed already at energies only slightly higher than so far accessible. Such a 
well-pronounced, step-like feature  would be a spectacular discovery 
in the next release of PAMELA data, or in future experiments like AMS-02.

In concluding, let us briefly 
address the consequences of this type of radiative corrections for possible 
dark matter induced contributions in other cosmic ray species. For 
\emph{neutrinos}, for the same reasons as discussed for secondary positrons, 
the only chance for large effects appears in situations with great enhancements 
of the annihilation into leptons, i.e. the channels $\mu^+\mu^-\gamma$ and 
$\tau^+\tau^-\gamma$. Still, just as in the case of positrons, these channels are unlikely to have a large impact on the \emph{total} neutrino yield.
A potentially 
interesting source for \emph{antiprotons}, on the other hand, are gluons from the annihilation into 
$t\bar t g$ final states -- which, in the stop coannihilation region, is 
considerably enhanced compared to the lowest order result in exactly the same way as 
the $t\bar t \gamma$ channel discussed in \cite{ibsusy}. However, since the 
lowest order annihilation rate is extremely small in the region of interest, 
the resulting absolute yield is still too small to be of great significance. 
This general expectation is in agreement with earlier studies 
\cite{Flores:1989ru}.

To summarize, we have shown that radiative corrections 
may significantly enhance the dark-matter induced positron yield and result in 
a pronounced spectral signature, a rising positron to electron ratio and 
a sharp cutoff in the positron spectrum at the 
neutralino mass $m_\chi$. To obtain such a spectral feature, similar to that 
observed by PAMELA \cite{pamela_idm08}, very large boost factors are 
needed. On the other hand, \emph{if} 
such a feature is observed, a strong enhancement can also be expected in the 
gamma-ray flux at photon energies close to $m_\chi$ \cite{ibsusy} (while the impact on, e.g., the expected antiproton spectrum would be negligible); such a 
cross-correlation would of course provide even stronger evidence for the dark 
matter nature of the signal. An unambiguous, testable prediction of this class of 
models is that 
the positron excess will be cut off at an energy not too far from the 
maximal energy presently reported by the PAMELA collaboration, 
as larger masses do not
reproduce the slope of the rising positron ratio (for a very similar situation 
for
another class of dark matter particles, see \cite{baltz}).

Finally, let us mention that the radiative 
corrections to the positron yield from neutralino annihilations that have been 
reported here have been implemented in the current, publically available version 5.0.1 of \ds\ \cite{ds}.

L.B.\ and J.E.\ thank the Swedish Research Council for support.


\end{document}